\documentclass[12pt]{article}
\usepackage{amsmath,amssymb}
\usepackage{epsfig}

\newcommand{\newc}{\newcommand}
\newc{\gev}{\hbox{\rm\,GeV}}
\newc{\tev}{\hbox{\rm\,TeV}}

\newc{\gsim}{\lower.7ex\hbox{$\;\stackrel{\textstyle>}{\sim}\;$}}
\newc{\lsim}{\lower.7ex\hbox{$\;\stackrel{\textstyle<}{\sim}\;$}}

\def\mpl{M_{\rm Pl}}
\def\order#1{{\cal O}(#1)}
\def\eq#1{eq.~(\ref{#1})}

\def\mgut{M_{\rm GUT}}

\def\vev#1{\langle {#1} \rangle}
\def\beq{\begin{equation}}
\def\eeq{\end{equation}}
\def\bea{\begin{eqnarray}}
\def\eea{\end{eqnarray}}

\begin{document}

\baselineskip=18pt

\setcounter{footnote}{0}
\setcounter{figure}{0}
\setcounter{table}{0}

\begin{titlepage}
\begin{flushright}
CERN-PH-TH/2007--089\\
LAPTH-1200/07
\end{flushright}
\vspace{.3in}

\vspace{.5cm}

\begin{center}

{\Large\sc{\bf Dynamical $\mu$ Term in Gauge Mediation}}

\vspace*{9mm}
\renewcommand{\thefootnote}{\arabic{footnote}}

\mbox{ {\sc A.~DELGADO}$^1$, {\sc G.~F.~GIUDICE}$^{\,1}$ and
{\sc P.~SLAVICH}$^{\,1,2}$}

\vspace*{0.9cm}

{\it $^1$ CERN, Theory Division,  CH--1211 Geneva 23, Switzerland.}

{\it $^2$ LAPTH, 9 Chemin de Bellevue, F--74941 Annecy-le-Vieux, France.}

\end{center}

\vspace{1cm}

\begin{abstract}
\medskip
\noindent
We address the $\mu$ problem of gauge mediation by considering a
singlet chiral superfield coupled to the Higgs and messenger
fields. We compute the soft terms generated below the messenger scale
and study the phenomenological consequences of the model. The
experimental bound on the Higgs mass provides a severe constraint that
identifies three special regions of parameters where the mass spectrum
and the collider signatures can be distinct from ordinary gauge
mediation.
\end{abstract}

\bigskip
\bigskip

\end{titlepage}


\section{The Origin of $\mu$ in Gauge Mediation}
\vspace*{-1mm}

Gauge mediation~\cite{dine1,dine2,dine3,grrev} is one of the most
appealing realizations of the supersymmetric extensions of the
Standard Model (SM), because of its high predictive power and
ultraviolet insensitivity. Indeed, the induced soft terms are
computable in terms of few parameters (generally three) and do not
lead to unacceptably large flavor violations, under mild assumptions
on the unknown interactions responsible for generating Yukawa
couplings.

There are two ingredients of gauge mediation that are still
obscure. One is the seed of supersymmetry breaking, which is expected
to have a dynamical origin, in order to explain naturally the
emergence of mass scales much smaller than the Planck mass $\mpl$. New
interactions have to transfer the original supersymmetry breaking from
the hidden sector to the messenger fields. Attempts to simplify this
structure and unify the hidden and messenger sectors have faced
various difficulties. One of the problems is that, once a hidden
sector with supersymmetry breaking is found, the couplings to the
messengers allow for new supersymmetric vacua. To address this
problem, models were constructed~\cite{meta,meta2} around metastable
false vacua where supersymmetry is broken, although the global minimum
remains supersymmetric. More recently, a new framework of theories
with these properties has been discovered~\cite{iss}, showing that
this approach is fairly general. This new development has revived the
interest in gauge mediation and led to the construction of many
interesting models~\cite{issmod}.

The second obscure ingredient is the seed of Peccei-Quinn (PQ)
symmetry breaking or, in other words, the origin of $\mu$, the
higgsino mass, and $B_\mu$, the square mass mixing the two scalar
Higgs doublets $H_d$ and $H_u$. The first aspect of this problem,
common to all supersymmetric models, is how to relate $\mu$ to the
soft masses of the other supersymmetric particles. This problem is
solved by assuming that the PQ symmetry is exact in the supersymmetric
limit, while $\mu$ is induced by supersymmetry-breaking
effects~\cite{giumas}. The second aspect appears only in theories with
computable soft terms, like gauge mediation, and it is expressed by
the generic prediction
\beq \frac{B_\mu}{\mu} \sim \frac FM ,
\label{murel}
\eeq 
where $F$ represents the value of the hidden-sector auxiliary field,
and $M$ is the mass of the mediating field.  Equation~(\ref{murel}) is
the consequence of a generic coupling in the K\"ahler potential
between the Higgs bilinear $H_d H_u$ and a spurionic superfield $\hat
X=1+\theta^2F/M$
\beq
\alpha \int d^4\theta  H_d H_u f(\hat X,{\hat X}^\dagger),
\label{opmu}
\eeq
where $\alpha$ represents the product of coupling constants and
possible loop factors required to generate the effective
interaction. The operator in \eq{opmu} simultaneously generates both
$\mu$ and $B_\mu$, leading to \eq{murel} independently of the actual
value of the loop-suppression factor $\alpha$. In gravity-mediated
models, $F/M$ corresponds to the natural scale of soft terms, and
\eq{murel} is fully satisfactory. However, in theories where the soft
terms are derived from $F/M$ through computable loop effects, like
gauge or anomaly~\cite{anom} or gaugino~\cite{gaug} mediation,
\eq{murel} predicts that the ratio $B_\mu /\mu$ is parametrically too
large, requiring an unnatural fine tuning.

One solution~\cite{dynmu} is to construct models where operators of
the form (\ref{opmu}) are absent, while couplings to the hidden sector
generate only the structure
\beq
 \int  d^4\theta H_d H_u D^2 f(\hat X,{\hat X}^\dagger),
\label{opmu2}
\eeq
where $D_\alpha$ is the supersymmetric covariant derivative. Since
$D^2 f(\hat X,{\hat X}^\dagger)$ is an antichiral superfield, the
operator in \eq{opmu2} generates $\mu$ but not $B_\mu$, which is then
induced at a higher order in perturbation theory. Other solutions use
other dynamical scales present in the hidden sector~\cite{scamu} or
required by the cancellation of the cosmological constant~\cite{cosmu}
to reproduce acceptable values of $\mu$ and $B_\mu$. It is also
possible to construct models with flavor symmetries~\cite{flamu} or an
R-symmetry~\cite{errmu}, leading to selection rules that invalidate
\eq{murel}.

An alternative approach is to introduce in the low-energy theory a new
SM singlet field $N$ coupled to the Higgs bilinear in the
superpotential
\beq
W=\lambda NH_d H_u -\frac{k}{3} N^3.
\label{supwn}
\eeq
A $Z_3$ symmetry forbids a bare $\mu$ term, and the coupling $k$ is
needed to break the global PQ symmetry. The effective $\mu$ and
$B_\mu$ terms can now be entirely generated by low-energy dynamics and
$\mu =\lambda \vev{N}$, $B_\mu =\lambda \vev{F_N} \sim \vev{N}^2$,
circumventing \eq{murel}.

In theories like gravity mediation, where $\mu$ is correctly generated
by supersymmetry breaking and \eq{murel} is successful, the
introduction of the singlet $N$ does not appear to be well
motivated. Not only is it superfluous, but it also introduces a
proliferation of new unknown parameters in the soft terms. Moreover, a
light singlet can potentially destabilize the hierarchy~\cite{destab},
as is the case when we embed the superpotential in \eq{supwn} into a
GUT. The situation is quite different in a theory with low
supersymmetry-breaking scale and computable soft terms, like gauge
mediation. In this case, the introduction of $N$ is essential to
bypass \eq{murel} and, in principle, it can be done at the price of
only two new parameters $\lambda$ and $k$ (to be compared with $\mu$
and $B_\mu$ of the minimal supersymmetric SM) in the full Lagrangian,
including soft terms. Moreover, as long as
$\sqrt{F}<10^8\gev$~\cite{grrev}, the coupling of $N$ to GUT fields
does not destabilize the hierarchy~\cite{stab}, and it can even be
used for the sliding-singlet mechanism~\cite{slid} to explain the
Higgs doublet-triplet splitting, in the limit $k\to 0$~\cite{pomar}.

The use of $N$ to generate the $\mu$ term in gauge mediation was
immediately suggested in the original paper on the
subject~\cite{dine1}, but it was also found that the specific form of
the soft terms in gauge mediation does not allow for a correct pattern
of electroweak breaking with an acceptable mass spectrum. The main
difficulty lies in generating a sufficiently large value of $\vev{N}$,
which requires either a negative soft square mass for the scalar field
$N$, or large A-terms for $\lambda$ and $k$ interactions. Neither of
these conditions can be satisfactorily obtained in gauge mediation,
unless one introduces new light fields coupled to $N$~\cite{dine1},
multiple singlets with appropriately adjusted couplings~\cite{dine2},
higher-dimensional interactions of $N$ with specific values of the
exponents~\cite{dine3}, or modifies the theory to include a new U(1)
gauge group, under which $N$ is charged, with new associated
fields~\cite{nir}. A thorough analysis of these possibilities has been
presented in ref.~\cite{muray}.

In ref.~\cite{wave} it was pointed out that a negative square mass for
$N$ and non-vanishing trilinears can be obtained if the singlet is
directly coupled to the messenger fields $\Phi$ in the superpotential
\beq
W=X\left(\kappa_1 \bar \Phi_1 \Phi_1 +  \kappa_2 \bar \Phi_2 \Phi_2\right) 
+\xi N \bar \Phi_1 \Phi_2,
\label{wnew}
\eeq
where $X$ is the hidden-sector superfield containing the Goldstino.
The form of \eq{wnew} can be guaranteed by symmetries, {\it e.g.} by a
discrete $Z_3$ with $Z_3[\Phi_1]= Z_3[\bar \Phi_2]=-1/3$,
$Z_3[\Phi_2]= Z_3[\bar \Phi_1]=Z_3[N]=1/3$, $Z_3[X]=0$, broken only at
the weak scale. The doubling of the
messenger field is necessary to avoid a kinetic mixing between $X$ and
$N$. Indeed, if both $X$ and $N$ coupled to the same bilinear $\bar
\Phi \Phi$, below the messenger mass $M$ we would find the one-loop
mixing in the effective K\"ahler potential
\beq
\frac{\xi d_\Phi}{16\pi^2} \int d^4\theta N X^\dagger \ln 
\left( \frac{XX^\dagger}{M^2}\right) +{\rm h.c.},
\eeq
where $d_\Phi$ is the dimensionality of the gauge representation of
$\Phi$. This generates a tadpole for the scalar field $N$ ($V_{\rm
eff} =(\xi d_\Phi /16\pi^2)NF^2/M$) that destabilizes the weak scale,
unless $\sqrt{F}<{\rm TeV}/\sqrt{\xi}$. Therefore the doubling of
messengers is necessary, unless we accept the unnatural case of an
extremely small value of $\xi$~\cite{han}. Instead of coupling the
singlet to both messengers as in eq.~(\ref{wnew}), one could also
consider coupling it to (the weak-doublet component of) just one
messenger, as in $N \bar\Phi_1 H_u$ or $N H_d \Phi_2$. This
alternative was discussed in ref.~\cite{chacko}.

In this paper we study the structure and the phenomenology of gauge
mediation with an extra singlet $N$ coupled to Higgs and messenger
fields as in eqs.~(\ref{supwn}) and (\ref{wnew}). This variation of
the ordinary SM extension with gauge-mediated supersymmetry breaking
(GMSB) will be referred to as the N-GMSB model. In section 2 we
compute the soft terms induced in the effective theory below the
messenger scale. In sections 3 and 4 we study the vacuum structure and
the phenomenology of N-GMSB. Our results are summarized in section 5.

\section{Generating the soft terms}

We now want to compute the soft terms of N-GMSB, the supersymmetric SM
with gauge mediation augmented by the superpotential interactions in
eqs.~(\ref{supwn}) and (\ref{wnew}). We start by treating
$X=M+\theta^2 F$ as a background non-dynamical field, and we will
later comment on the case in which $X$ can propagate. We also assume
that $\Phi$ ($\bar \Phi$) belongs to a fundamental (antifundamental)
of $SU(5)$ and we introduce separate couplings for the interactions of
the $SU(2)$ doublet ($\Phi^D$) and the $SU(3)$ triplet ($\Phi^T$)
contained in the messenger multiplet, expanding the relevant terms in
the superpotential as
\beq 
W=X\sum_{i=1}^2 \left( \kappa_i^D {\bar \Phi}_i^D \Phi_i^D +
\kappa_i^T {\bar \Phi}_i^T \Phi_i^T \right)
+N \left( \xi_D {\bar \Phi}_1^D \Phi_2^D 
+ \xi_T {\bar \Phi}_1^T \Phi_2^T \right) 
+\lambda N H_d H_u -\frac{k}{3} N^3.
\label{supp}
\eeq

We define the supersymmetry-breaking mass and interaction terms for
the Higgs bosons of the effective theory valid below the messenger
mass as
\beq 
V_{\rm soft}= {\tilde m}_{H_u}^2 |H_u|^2+{\tilde m}_{H_d}^2
|H_d|^2+{\tilde m}_{N}^2 |N|^2+\left( \lambda A_\lambda NH_d H_u -\frac
k3 A_k N^3 +{\rm h.c.} \right) .
\label{vsoft}
\eeq
Soft terms are obtained by integrating out the messengers at one loop,
for $A$ terms, and at two loops, for scalar masses. Instead of
computing the full set of diagrams, a rather daunting task, we use the
method proposed in refs.~\cite{wave,wave2} to extract
supersymmetry-breaking effects from wave-function renormalization.

We first write the one-loop Renormalization Group (RG) equations for
the field wave-function renormalizations $Z_\alpha$ and the coupling
constants $\lambda_i$ as
\beq 
\frac{d \ln Z_\alpha}{d\ln Q} =\gamma_\alpha,~~~~~\alpha=N,H_d,H_u
\label{zrun}
\eeq
\beq
\frac{d \lambda_i^2}{d\ln Q} =\beta_{\lambda_i} ,~~~~~ 
\lambda_i=\lambda ,k,\xi_D,\xi_T,g,g^\prime ,g_s  .
\label{betarun}
\eeq
Here $Q$ is the renormalization scale, $\gamma_\alpha$ are the
anomalous dimensions, and $\beta_{\lambda_i}$ are the beta functions.  Retaining
only the leading terms of an expansion in powers of $F/M$, the soft
supersymmetry-breaking parameters in \eq{vsoft} are given by (see
appendix B)
\beq 
{\tilde m}_{H_u}^2={\tilde m}_{H_d}^2= -{\cal Z}_H^{\prime
\prime} \frac{F^2}{M^2},~~~~ {\tilde m}_{N}^2=-{\cal Z}_N^{\prime
\prime} \frac{F^2}{M^2},
\label{sf1}
\eeq
\beq 
A_\lambda =\left( {\cal Z}_N^{\prime} +2 {\cal Z}_H^{\prime}
\right) \frac FM ,~~~~ A_k =3 {\cal Z}_N^{\prime} \frac FM ,
\label{sf2}
\eeq
\beq 
\left. {\cal Z}_\alpha^{\prime}\right|_{Q=M} = 
 \frac{\Delta\gamma_\alpha}{2},
~~~~ \left. {\cal Z}_\alpha^{\prime\prime}\right|_{Q=M} =
\frac 14\sum_i \left[ \beta_{\lambda_i}^{(+)}
\frac{\partial \left( \Delta \gamma_\alpha \right)}{\partial \lambda_i^2}
-\Delta \beta_{\lambda_i} 
\frac{\partial  \gamma_\alpha^{(-)}}{\partial \lambda_i^2}
\right]_{Q=M} ,
\label{derivv}
\eeq
where we have defined $\Delta X\equiv [X^{(+)}-X^{(-)}]_{Q=M}\,$ (with
$X=\beta_{\lambda_i}, \gamma_\alpha$) as the discontinuity at the messenger
scale, and $X^{(\pm)}$ are the values of $X$ in the theory above and
below $M$, respectively.  ${\cal Z}_\alpha^{\prime}$ is proportional
to the discontinuity of the anomalous dimension at the messenger scale
$M$, and ${\cal Z}_\alpha^{\prime\prime}$ depends on a combination of
the discontinuities of the anomalous dimension and
beta-functions. Such discontinuities can be present if some particles
contributing to $\beta_{\lambda_i}$ or $\gamma$ are integrated out at the scale
$M$.

To obtain explicit formulae for the matching conditions on the soft
terms at the scale $M$ we only need to specify the anomalous
dimensions, beta functions and their discontinuities at $M$. This can
be easily done from the RG equations given in appendix A. The
anomalous dimensions and their discontinuities are given by
\beq 
\gamma_{H_u,H_d}^{(-)} = -\frac{1}{16\pi^2} \left( 2\lambda^2
-3g^2-g^{\prime 2} +{\rm ``Yukawa"}\right) ,~~~~
\Delta \gamma_{H_u,H_d} =0\,,
\label{andim1}
\eeq 
\beq 
\gamma_{N}^{(-)} = -\frac{1}{8\pi^2} \left( 2\lambda^2+2 k^2 \right),~~~~
\Delta \gamma_N =-\frac{1}{8\pi^2} \left( 2\xi_D^2
+3\xi_T^2\right) .
\label{andim2}
\eeq
In \eq{andim1}, we have not specified the ``Yukawa" contribution,
which is different for $H_d$ and $H_u$, since it does not lead to any
discontinuity at the scale $M$ and therefore does not contribute to
soft masses up to two-loop order.  The beta-functions and their
non-vanishing discontinuities that contribute to ${\cal
Z}_\alpha^{\prime\prime}$ in \eq{derivv} are
\bea
\label{discont}
&& \beta^{(+)}_{\xi_D}=\frac{\xi_D^2}{8\pi^2}\,
\biggr( 2\lambda^2 +2k^2 +4 \xi_D^2+3\xi_T^2-3g^2-g^{\prime 2}\biggr)~,
\nonumber \\ &&
\beta^{(+)}_{\xi_T}=\frac{\xi_T^2}{8\pi^2}
\left( 2\lambda^2 +2k^2 +2 \xi_D^2+5\xi_T^2-\frac{16}{3} 
g_s^2-\frac 49 g^{\prime 2}\right) ~,
\nonumber \\ &&
\Delta \beta_\lambda =\frac{\lambda^2}{8\pi^2} \left( 2\xi_D^2
+3\xi_T^2\right) ,~~~ \Delta \beta_k =\frac{3k^2}{8\pi^2} \left(
2\xi_D^2 +3\xi_T^2\right) ,~~~ \Delta
\beta_{g_i}=c_i\,n\,\frac{g_i^4}{8\pi^2}~.
\eea
Here the gauge couplings $g_i$ are ordered as $(g^\prime,g,g_s)$ and
the constants $c_i$ are $(5/3,1,1)$; $n$ is the number of messenger
pairs (we take $n=2$). Finally, the matching conditions on the soft
terms at the scale $M$ are explicitly written as
\beq
A_\lambda =\frac{A_k}{3} =-\frac{1}{16\pi^2}\,
\left( 2\xi_D^2+3\xi_T^2 \right) \frac{F}{M}\,,
\label{aterm}
\eeq
\beq 
\label{mhud}
{\tilde m}_{H_u}^2={\tilde m}_{H_d}^2=\frac{1}{(16\pi^2)^2}\,
\left[\, n\left(
\frac{3g^4}{2}+\frac{5g^{\prime 4}}{6}\right) - \lambda^2\left(2
\xi_D^2+3\xi_T^2\right)  \right]\, \frac{F^2}{M^2} \,,
\eeq
\bea
{\tilde m}_{N}^2&=&\frac{1}{(16\pi^2)^2}\,
\left[ 8\xi_D^4
  +15\xi_T^4+12\xi_D^2\xi_T^2 -
  16g_s^2\xi_T^2 -6g^2\xi_D^2 - 2g^{\prime 2}
  \left( \xi_D^2 +\frac{2}{3} \xi_T^2 \right) 
  \right. \nonumber \\
&& 
 - 4k^2 \left( 2\xi_D^2+3\xi_T^2\right)\biggr]\,
  \frac{F^2}{M^2}\,.
\label{spot}
\eea
Here all couplings and parameters are evaluated at $Q=M$. 

Note that there is no one-loop contribution to ${\tilde m}_N^2$ of
order $F^2/M^2$.  Indeed, the messenger interactions in \eq{supp} are
invariant under independent chiral reparametrizations of the fields
$N$ and $X$ (with messenger fields transforming appropriately),
therefore the one-loop K\"ahler potential must be of the form $\int
d^4 \theta N^\dagger N \ln X^\dagger X$ and cannot induce a soft mass
for $N$.

A one-loop contribution to ${\tilde m}_N^2$ can be generated only at
higher orders in the $F/M^2$ expansion, and we find
\beq
{\tilde m}_N^2 = -\frac{2\,\xi_D^2+3\,\xi_T^2}{16\pi^2}\, 
 \frac{F^4}{\kappa_1^2M^6} \,f\left( \frac{\kappa_2^2}{\kappa_1^2}\right)
 ~+~{\cal O}\left( \frac{F^6}{M^{10}}\right) ,
\eeq
\beq
f(x)=\frac{1-x^2+2\,x\ln x}{(1-x)^3} .
\eeq
This contribution is always negative. However, it is negligible with
respect to the one in \eq{spot}, as long as $M\gsim 4\pi F/M\simeq
10^3$~TeV. As we will see in section \ref{secnum}, this condition is
satisfied in our study, since the Higgs mass bound selects large
values of $M$. A variant to induce a one-loop contribution to ${\tilde
m}_N^2$ of order $F^2/M^2$ is given by models with several
hidden-sector fields with non-vanishing vacuum expectation values
(vevs). This case can be parametrized by a superpotential interaction
$X_1\bar \Phi_1 \Phi_1 + X_2\bar \Phi_2 \Phi_2$, with
$X_{1,2}=M_{1,2}+\theta^2 F_{1,2}$ where $M_{1,2}$ and $F_{1,2}$ are
independent. Now we have to consider two different messenger
thresholds and the one-loop contribution is given by
\beq
{\tilde m}_N^2 = \frac{2\,\xi_D^2+3\,\xi_T^2}{16\pi^2}\,
\left( \frac{F_1}{M_1}-\frac{F_2}{M_2}\right)^{\!2} \, 
g\left( \frac{M_2^2}{M_1^2}\right) ,
\label{twof}
\eeq
\beq
g(x)=\frac{x}{(x-1)^3}\left[ 2\,(1-x) +(1+x)\ln x\right] .
\eeq
Note that the contribution in \eq{twof} is always positive, and
vanishes when the supersymmetry breaking is universal, $F_1/M_1 =
F_2/M_2$, (as is the case for a single $X$ field, considered in this
paper, where the coupling constants $\kappa_{1,2}$ drop out from the
ratio $F_{1,2}/M_{1,2}$) or when one messenger threshold decouples
($M_1$ or $M_2 \to \infty$).

The matching conditions at the messenger scale on the soft masses for
gauginos, squarks and sleptons are given by the usual expressions of
gauge mediation
\beq
\label{gaugino}
M_i = n\,c_i\,\frac{\alpha_i}{4\,\pi}\,\frac FM\,,
\eeq
\beq
\label{sfermions}
m^2_{\tilde f} = 2 \,n\,\sum_i\,c_i\,C_i^{\tilde f}\, 
\frac{\alpha^2_i}{(4\,\pi)^2}\,\frac{F^2}{M^2}\,,
\eeq
where the coefficients $c_i$ and $n$ are given below
eq.~(\ref{discont}), and $C_i^{\tilde f}$ is the quadratic Casimir
invariant for the scalar $\tilde f$ under the gauge group with
coupling $\alpha_i$. The matching conditions at the messenger scale on
the trilinear $A$-terms corresponding to Yukawa interactions vanish at
leading order, while nonzero values are generated at the weak scale by
RG evolution.
 
As is well known, the couplings $\kappa_{1,2}$ of $X$ to the messenger
fields do not affect the soft terms, since they drop out of the ratio
$F/M$. However, when messengers are coupled to $N$, a propagating $X$
field gives a two-loop diagram that contributes to ${\tilde
m}_{N}^2$. Indeed, with a dynamical $X$ we obtain an extra
contribution to $\beta_{\xi_{D,T}}^{(+)}$ giving
\beq 
\delta \beta_{\xi_{D,T}}^{(+)}= -\frac{\xi_{D,T}^2}{8\pi^2} \left(
{\kappa_1^{D,T}}^2 + {\kappa_2^{D,T}}^2 \right) .  
\eeq
This leads to an extra term to be added to \eq{spot},
\beq 
\delta {\tilde m}_{N}^2 =\frac{1}{(16\pi^2)^2}\,
\left[ 2\,\xi_D^2 \left( {\kappa_1^{D}}^2
+ {\kappa_2^{D}}^2 \right) + 3\,\xi_T^2 \left( {\kappa_1^{T}}^2 +
{\kappa_2^{T}}^2 \right) \right] \,\frac{F^2}{M^2}.
\label{sloppa}
\eeq
In the rest of the paper we will restrict our analysis to the case in
which $X$ is a spurion representing only the mass parameters $M$ and
$F$ (or to the case in which $X$ propagates, but $\kappa_i^{D,T}$ are
negligible with respect to the other coupling constants) and neglect
the contribution in \eq{sloppa}.

\section{Vacuum Structure and Higgs Boson Masses}

To determine the mass spectrum of the low-energy limit of N-GMSB,
which essentially can be viewed as a constrained version of the
Next-to-Minimal Supersymmetric Standard Model (N-MSSM), we must
compute all the Lagrangian parameters at some renormalization scale of
the order of the weak scale, where we impose the minimization
conditions of the Higgs potential. The model has five unknown input
parameters: the singlet couplings $\lambda$ and $k$; the messenger
mass $M$; the effective supersymmetry breaking scale $F/M$; the
unified value $\xi_{U} \equiv \xi_{T,D}(\mgut)$ for the
singlet-messenger couplings at the GUT scale (defined as the scale
where the couplings $g$ and $\sqrt{5/3}\,g^{\prime}$ meet). Other
required inputs are the gauge and third-family Yukawa couplings, which
we extract at a low reference scale equal to the pole top mass $M_t =
170.9$ GeV \cite{mtop} from the known values \cite{pdg} of the fermion
masses and of the SM input parameters $G_F,\,M_Z,\,\sin^2\theta_W$ and
$\alpha_s(M_Z)\,$. We use tree-level formulae for the determination of
all the couplings but the top Yukawa coupling $h_t$, for which we
include one-loop corrections. The soft supersymmetry-breaking masses
and interaction terms for Higgs bosons, gauginos and sfermions are
determined at the messenger scale $M$ by means of
eqs.~(\ref{aterm})--(\ref{spot}) and
eqs.~(\ref{gaugino})--(\ref{sfermions}). Finally, we determine all the
parameters of the N-GMSB Lagrangian at a renormalization scale $M_S$
that, in order to minimize the dominant $\order{h_t^4}$ one-loop
corrections to the Higgs potential, we choose as the geometric average
of the two stop masses, i.e.~$M_S = \sqrt{m_{\tilde t_1}m_{\tilde
t_2}}$. To this purpose we use the RG equations of the effective
theories valid between the different mass scales: SM between $M_t$ and
$M_S$; N-MSSM between $M_S$ and $M$; N-GMSB including the messenger
sector\footnote{We neglect possible self-interactions in the hidden
sector. Otherwise, as argued in ref.~\cite{schmaltz}, we should
consider one more effective theory valid between the scales $M$ and
$\sqrt{F}$.} between $M$ and $\mgut$.  The explicit formulae for the
RG equations are given in the appendix A.  Since the boundary
conditions on the various parameters are given at different
renormalization scales, and some of them depend on the vevs
$\vev{H_u}$ and $\vev{H_d}$ determined by the minimization of the
Higgs potential, we need to iterate the procedure until it converges.

The tree-level scalar potential along the neutral components of the
fields $H_{d,u}$ and $N$ is
\bea
\label{vtree}
V_0&=&\left| \lambda H_dH_u -k N^2\right|^2 +\lambda^2 \left| N\right|^2 
\left(   \left| H_d\right|^2 +  \left| H_u\right|^2\right) 
+\frac{g^2+g^{\prime 2}}{8}\left(   \left| H_d\right|^2 -  
\left| H_u\right|^2\right)^2 \nonumber \\
&+&\left( \lambda A_\lambda NH_dH_u -\frac{k}{3}A_k N^3 +{\rm h.c.} \right) + 
{\tilde m}_{H_u}^2 |H_u|^2+{\tilde m}_{H_d}^2
|H_d|^2+{\tilde m}_{N}^2 |N|^2.
\eea
The minimization conditions of the scalar potential with respect to
the three Higgs fields allow us to determine the vevs $\vev{H_u}\,,
\vev{H_d}$ and $\vev{N}$. In practice, we treat the electroweak
symmetry-breaking scale $v^2 \equiv \vev{H_u}^2 + \vev{H_d}^2 \approx
(174 \gev)^2$ as an input parameter extracted at $Q=M_t$ from the
Fermi constant $G_F$ and evolved up to $Q=M_S$ with the SM RG
equations. The minimization conditions can therefore be used to
determine one of the unknown input parameters, reducing their number
to four. In terms of parameters computed at the scale $M_S$, the
minimization conditions can be expressed as
\bea
\mu^2 &=& \frac{{\tilde m}_{H_d}^2-{\tilde m}_{H_u}^2\tan^2\beta}
{\tan^2\beta -1}-\frac{g^2+g^{\prime\,2}}{4}\,v^2~, \label{min1}\\ 
&&\nonumber\\
\sin 2\beta&=& \frac{2\,B_\mu }{
{\tilde m}_{H_d}^2+{\tilde m}_{H_u}^2+2\mu^2},
\label{min2}\\
&&\nonumber\\
2\,\frac{k^2}{\lambda^2}\, \mu^2 -\frac k\lambda A_k \,\mu + {\tilde m}_{N}^2
&=&
\lambda^2v^2 \left[ -1+\left( \frac{B_\mu}{\mu^2}+\frac k\lambda \right) 
\frac{\sin 2\beta}{2} +\frac{\lambda^2\,v^2\,\sin^2 2\beta}{4\,\mu^2}  
\right]~,\label{min3}
\eea
where, to highlight the analogy between
eqs.~(\ref{min1})--(\ref{min2}) and the corresponding minimization
conditions in the usual MSSM, we define $\tan\beta \equiv
\vev{H_u}/\vev{H_d}$ and introduce the quantities $\mu$ and $B_\mu\,$:
\beq
\label{muBmu}
\mu~\equiv ~\lambda\, \vev{N}~,~~~~~~~~
B_\mu~\equiv~ \frac{k}{\lambda}\,\mu^2 -A_\lambda\,\mu 
-\frac{\lambda^2\, v^2}{2}\,\sin 2\beta~.
\eeq
Eqs.~(\ref{min1})--(\ref{min3}) depend non-trivially on the various
parameters and must be solved numerically. They provide us with values
at $Q=M_S$ for $\vev{N},\, \tan\beta$ and a third parameter that we
choose to be $k$. The remaining input parameters are thus $M,\, F/M,\,
\xi_{U}$ and $\lambda$ (the latter given at the scale $M_S$). Without
loss of generality, we can take $\lambda$ real and positive and
exploit the freedom to redefine the phases of the fields to make sure
that $\tan\beta$ is also positive; this involves flipping the signs of
$\vev{N}$ and $k$ if the numerical solution of the minimum equations
gives a negative value for $\tan\beta$. We also choose a basis in
which $F/M$ is real and positive, so that the gaugino masses in
eq.~(\ref{gaugino}) are positive. In the general N-MSSM, the relative
phases between the $A$-terms and the gaugino masses cannot be removed
and are physical sources of CP violation. However, in the N-GMSB these
phases are all zero and, in the field basis we have chosen,
$A_\lambda$ and $A_k$ at the messenger scale turn out to be real and
negative, see eq.~(\ref{aterm}).

Before discussing our treatment of the radiative corrections in the
N-MSSM Higgs sector and moving on to the numerical analysis, we
present some analytical considerations that help understanding the
vacuum structure of the theory. After LEP unsuccessful searches for
the Higgs boson and for new particles, supersymmetric models suffer
from a mild fine-tuning problem that requires a certain separation of
scales between $v$ and the superparticle masses. Therefore the only
acceptable region of parameters has to lie very close to the
``critical line" separating the phases with broken and unbroken
electroweak symmetry~\cite{crit}. In practice, this means that we can
find $\vev{N}$ by setting $v=0$ in \eq{min3}, and then imposing the
critical condition for electroweak breaking on the effective Higgs
potential at fixed $N$ background value. From \eq{min3} we obtain
\beq
\mu = \frac{\lambda}{k}\, A_k \,w +\order{v^2}~,
~~~~~w\equiv\frac{1+\sqrt{1-8z}}{4}~,
~~~~~z\equiv \frac{{\tilde m}_{N}^2}{A_k^2}~.
\label{trovmu}
\eeq
The non-trivial vacuum for $N$, corresponding to \eq{trovmu}, exists
only for $z<1/8$, but we have to impose $z<1/9$ to insure that this
vacuum is deeper than the origin $\vev{N}=0$. This condition then
leads to $w> 1/3$.

Equations (\ref{min1}) and (\ref{min2}), in the limit $v \to 0$,
correspond to the critical condition that the origin of the effective
Higgs potential with $\vev{N}$ fixed has locally one flat direction
and non-negative second derivatives:
\beq
\left( {\tilde m}_{H_d}^2 +\mu^2\right)  
\left( {\tilde m}_{H_u}^2 +\mu^2\right) =B_\mu^2~,
\label{min4}
\eeq
\beq
{\tilde m}_{H_d}^2+{\tilde m}_{H_u}^2+2\mu^2 >0~.
\label{min5}
\eeq
Combining eqs.~(\ref{muBmu})--(\ref{min4}) and neglecting terms of
$\order{v^2}$ we obtain the critical line in the $\lambda$--$k$ plane:
\beq
\frac{\lambda^2}{k^2}=\frac{\left( A_k\,w-A_\lambda\right)^2}
{A_k^2\,w^2+\frac{k^2}{\lambda^2}\,{\tilde m}_{H_d}^2}
-\frac{{\tilde m}_{H_u}^2}{A_k^2\,w^2}~.
\label{critline}
\eeq
Equation~(\ref{critline}) shows that the critical line gives an
approximately linear relation between $\lambda$ and $k$, distorted
only by the small ${\tilde m}_{H_d}^2$ contribution and by RG effects.
At large values of $\lambda$ (and $k$), the critical line is
interrupted either by perturbative constraints on $\lambda$ and the
top Yukawa coupling, or by the appearance of a minimum with
$\vev{H_u}\ne 0$ and $\vev{H_d}=\vev{N}=0$, which becomes deeper than
the correct vacuum unless
\beq
\label{wrongvac}
k^2<\frac{(g^2+g^{\prime 2})\,A_k^4}{2 \,{\tilde m}_{H_u}^4}\,w^3
\left( w-\frac 13 \right) .
\eeq 

It is also interesting to note that the combination of the
minimization conditions of the scalar potential with the boundary
conditions on the Higgs trilinear couplings leads to a definite
prediction for the sign\,\footnote{\,Our choice for the sign of the
superpotential term $\lambda N H_d H_u$ in eq.~(\ref{supp})
corresponds to the convention in which the off-diagonal element of the
stop mass matrix contains $m_t\,\mu\,\cot\beta$ and the chargino mass
matrix contains $-\mu$~.} of $\mu$. Indeed, it can be seen from
eqs.~(\ref{min2}) and (\ref{min5}) that the condition $\tan\beta>0$
requires $B_\mu$ to be positive. Combining eqs.~(\ref{muBmu}) and
(\ref{trovmu}), and neglecting terms of $\order{v^2}$, one gets
\beq
\label{condmu}
B_\mu \simeq \mu\,\left(A_k\,w - A_\lambda\right)~.
\eeq
At the messenger scale $A_k$ is negative and equal to $3\, A_\lambda$,
see eq.~(\ref{aterm}). If the effect of the RG evolution of the soft
supersymmetry-breaking parameters down to the scale $M_S$ is
neglected, the condition $w>1/3$ constrains $\mu$ to be always
negative. In practice we find that, even though the RG evolution can
alter the relation between $A_k$ and $A_\lambda$, all the
phenomenologically viable solutions to the minimization conditions of
the scalar potential have indeed $\mu<0$.

In the limit $\vev{N} \gg v$ the tree-level squared masses of the two
CP-odd and three CP-even neutral scalars are
\beq
m_{a_1}^2 ~ = ~ \frac{\mu^2+{\tilde m}_{H_d}^2}{\sin^2\beta}~+ 
~ \order{v^2},~~~~~~~~~
m_{a_2}^2 ~ = ~ \frac{3}{w} \left(\frac k\lambda \,\mu \right)^2~+ 
~ \order{v^2},
\eeq
\beq
\label{massh1}
m_{h_1}^2 ~=~  M_Z^2\,\cos^22\beta + \lambda^2\,v^2\,\left\{
\sin^22\beta - \frac{\left[ \frac \lambda k 
+\left(\frac{A_\lambda}{2wA_k}-1\right)
\,\sin2\beta\right]^2}
{1-\frac{1}{4w}}\right\}~+ ~ \order{v^4},
\eeq
\beq
m_{h_2}^2  ~=~ m_{a_1}^2 ~ + ~ \order{v^2},~~~~~~~~~
m_{h_3}^2 ~=~ \frac{4w-1}{3}   ~m_{a_2}^2 ~ + ~ \order{v^2}~.
\eeq
Analogously to the decoupling limit of the usual MSSM, the lightest
CP-even Higgs boson $h_1$ has SM-like couplings to fermions and gauge
bosons, and its mass is of $\order{v^2}$, with an additional
contribution -- in curly brackets in eq.~(\ref{massh1}) -- that is not
present in the MSSM; the condition $w>1/3$ also ensures that the
second term in the curly brackets is always negative. The CP-even
boson $h_2$ and the CP-odd boson $a_1$ are heavy and have couplings
similar to those of the MSSM Higgs bosons $H$ and $A$ (the same
applies to the charged boson). The CP-even boson $h_3$ and the CP-odd
boson $a_2$ are mostly singlet and are mostly decoupled from matter
fields. We find $m_{h_3}> m_{a_2}$ for $z<-1$, and $m_{h_3}< m_{a_2}$
for $-1<z<1/9$.

Detailed studies of the N-MSSM Higgs sector date back to the nineties
\cite{nmssmhiggs}. It is also well known that in supersymmetric models
the radiative corrections involving top and stop loops can give a
substantial contribution to the Higgs boson masses \cite{mssmcorr},
and they must be taken into account for a meaningful comparison with
the mass bounds from direct searches at LEP \cite{LEPhiggs}. The
dominant one-loop corrections, enhanced by four powers of the top
Yukawa coupling $h_t$, can be computed in the effective potential
approach. The radiatively corrected effective potential for the Higgs
fields can be written as $V_{\rm eff} = V_0 + \Delta V$, where $V_0$
is given in eq.~(\ref{vtree}) and the correction $\Delta V$ is
expressed in terms of field-dependent masses and mixing angles. The
radiative corrections to the minimization conditions of the scalar
potential are taken into account by replacing in
eqs.~(\ref{min1})--(\ref{min3})
\beq
\label{tadpoles}
~~~~~~~~~~ \tilde m^2_{\phi_i} ~ \longrightarrow ~ 
\tilde m^2_{\phi_i} ~+~ \frac{1}{\vev{\phi_i}}\,\left.
\frac{\partial \,\Delta V}{\partial \,\phi_i}\right|_{\rm min}
~~~~~~~~~~ \phi_i = (H_d,\,H_u,\, N)~,
\eeq
where the subscript ``min'' means that the Higgs fields are set to
their vev after computing the derivative of the potential. The
radiative corrections to the $3\!\times\!3$ mass matrices for the
CP-even and CP-odd Higgs bosons are in turn:
\beq
\label{masscorr}
\left(\Delta{\cal M}_S^2\right)_{ij} = \frac{1}{2}\,\left.
\frac{\partial^2 \,\Delta V}{\partial \,{\rm Re}\, \phi_i\,\partial 
\,{\rm Re}\, \phi_j}\,
\right|_{\rm min}~,~~~~~~~~
\left(\Delta{\cal M}_P^2\right)_{ij} = \frac{1}{2}\,\left.
\frac{\partial^2 \,\Delta V}{\partial \,{\rm Im}\, \phi_i\,\partial 
\,{\rm Im}\, \phi_j}\,
\right|_{\rm min}~.
\eeq

We have explicitly computed the dominant $\order{h_t^4}$ corrections
to the minimization conditions and to the Higgs mass matrices given in
eqs.~(\ref{tadpoles}) and (\ref{masscorr}), and checked that our
results agree with those available in the literature \cite{nmssmcorr}.
In addition, we include in our determination of the Higgs masses the
one-loop leading logarithmic corrections of $\order{h_t^2 g^2,\,
h_t^2\lambda^2}$ (by multiplying the mass matrices by appropriate
wave-function-renormalization factors) and the two-loop leading
logarithmic corrections of $\order{h_t^4 g_s^2,\, h_t^6}$. Finally,
after diagonalizing the Higgs mass matrices we include the one-loop
leading logarithmic corrections of $\order{\lambda^4}$ to the mass of
the lightest CP-even Higgs boson $h_1$, computed in the limit where
$\vev{N} \gg v$. These corrections are accounted for by the term
\beq
\label{extracorr}
\Delta m_{h_1}^2 = 
-\frac{3 \,\lambda_{h_1}^2\,v^2}{4\,\pi^2}\,\ln\frac{Q^2}{M_t^2}~,
\eeq
where the lightest-Higgs quartic coupling is defined as $\lambda_{h_1}
= m_{h_1}^2/(2\,v^2)$, and the tree-level mass of $h_1$ in the limit
$\vev{N} \gg v $ was given in eq.~(\ref{massh1}).  The correction
$\Delta m_{h_1}^2$ in eq.~(\ref{extracorr}) can be numerically
relevant only if $\lambda$ is fairly large. It also includes some (not
all) of the one-loop leading logarithmic corrections that involve the
electroweak couplings, but the contribution of such terms is generally
small.

We have compared the results of our calculation of the Higgs boson
masses with those of the general N-MSSM model using the public
computer code {\tt NMHDECAY} \cite{nmhdecay}, which includes also the
corrections controlled by the bottom Yukawa coupling as well as a more
refined treatment of the one-loop leading logarithmic corrections
controlled by powers of $\lambda$ and of the electroweak couplings. We
find that, in points of the parameter space that will be relevant to
our analysis, the two determinations of the lightest CP-even Higgs
mass $m_{h_1}$ agree within 5 GeV, with {\tt NMHDECAY} predicting in
general smaller values of $m_{h_1}$ than our calculation. We consider
this agreement satisfactory, given the approximations involved in our
calculation -- we neglect the one-loop electroweak corrections and the
two-loop non-leading-logarithmic terms of $\order{h_t^4 g_s^2,\,
h_t^6}$ -- and the unavoidable uncertainty coming from uncomputed
higher-order corrections.

\section{Phenomenology}
\label{secnum}

As discussed in the previous section, the requirement of successful
breaking of the electroweak symmetry reduces the free parameters of
the model to four: the messenger mass $M$, the effective
supersymmetry-breaking scale $F/M$, the GUT-scale singlet-messenger
coupling $\xi_{U}$ and the singlet-Higgs coupling $\lambda$ computed
at a renormalization scale $M_S$ of the order of the average stop
mass. The parameters $M$ and $F/M$ should be chosen in such a way that
the $\order{h_t^4}$ radiative corrections involving top and stop loops
are large enough to lift $m_{h_1}$ above the bound from direct
searches at LEP (large values of the supersymmetric scale also imply
that the heavy Higgs bosons are essentially decoupled, thus the LEP
lower bound of 114.4 GeV \cite{LEPhiggs} on the mass of a SM-like
Higgs boson applies). In the usual GMSB the condition that the
trilinear Higgs-stop coupling $A_t$ be zero at the messenger scale
results in a small stop mixing at the weak scale. Therefore, a large
value of $M_S$, greater than a (few) TeV, is required to make
$m_{h_1}$ large enough.  In the model with an additional singlet, on
the other hand, positive contributions to $m_{h_1}$ can arise when
$\lambda$ is large and $\tan\beta$ is small, see
eq.~(\ref{massh1}). However, the conditions of correct electroweak
symmetry breaking and perturbativity of the couplings up to the GUT
scale require $\lambda(M_S) \,\lsim \,0.55$, and a sizeable
contribution to $m_{h_1}$ from radiative corrections remains
necessary. We will therefore choose a value of $F/M$ large enough to
result in an average stop mass of the order of 2 TeV. A large
messenger mass $M$ is also required to ensure that a sizeable value of
$A_t$ is generated by the RG evolution down to the weak scale.

\begin{figure}[p]
\vspace*{-5mm}
\begin{center}
{\epsfig{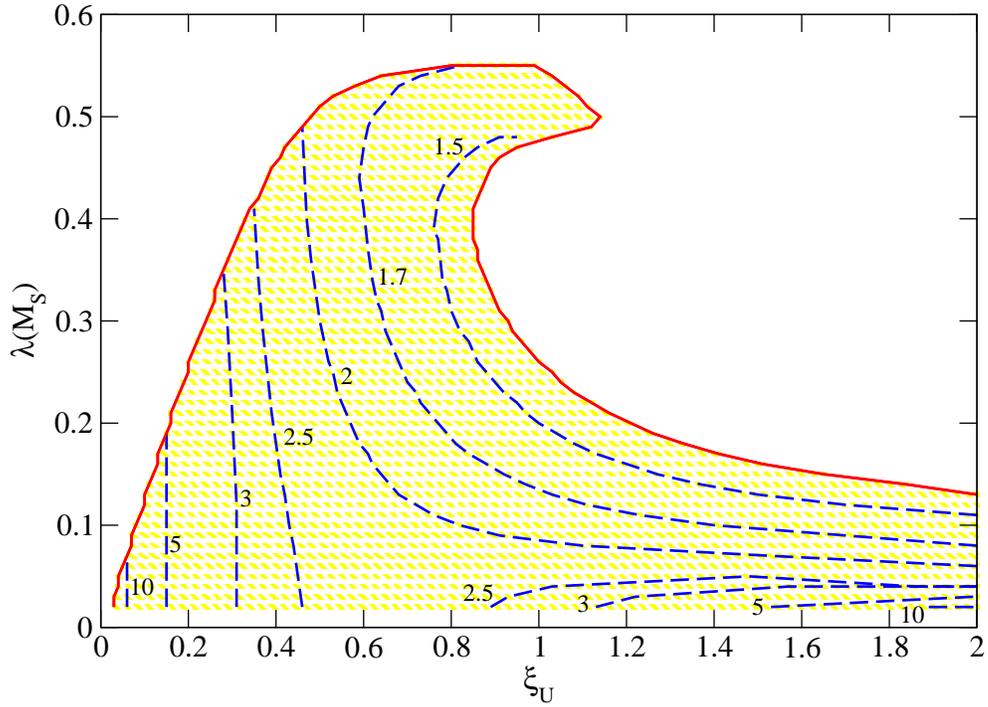}}
\end{center}
\vspace*{-5mm}
\caption{Values of $\tan\beta$ in the $\xi_U-\lambda(M_S)$ plane, for
$M = 10^{13}$ GeV and $F/M = 1.72 \times 10^5$ GeV.}
\label{fig-laxitb}
\end{figure}

\begin{figure}[p]
\begin{center}
{\epsfig{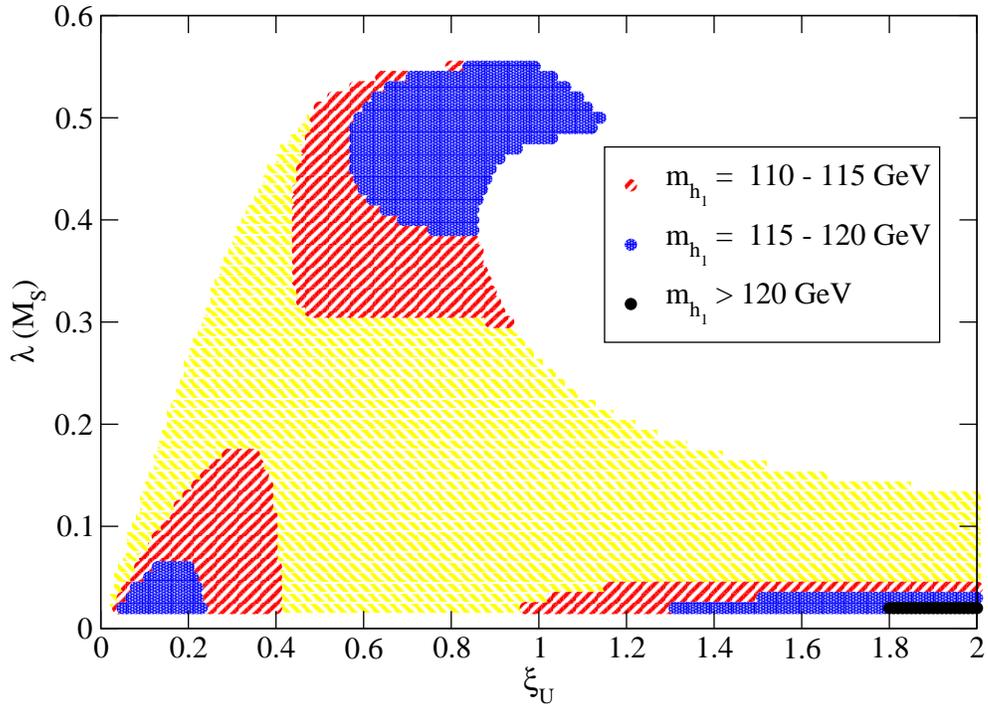}}
\end{center}
\vspace*{-5mm}
\caption{Mass of the lightest CP-even Higgs boson $h_1$ in the
$\xi_U-\lambda(M_S)$ plane, for
$M = 10^{13}$ GeV and $F/M = 1.72 \times 10^5$ GeV.}
\label{fig-laxi}
\end{figure}

Figure \ref{fig-laxitb} shows the values of $\tan\beta$ that result
from the minimization of the scalar potential over the plane
$\xi_U-\lambda(M_S)$.  We choose $M = 10^{13}$ GeV and $F/M = 1.72
\times 10^5$ GeV, resulting in an average stop mass $M_S =
\sqrt{m_{\tilde t_1}m_{\tilde t_2}}$ that varies between 1.9 and 2.1
TeV. Outside the allowed region, which is delimited by the solid (red)
line, no satisfactory solution to the minimization conditions of the
scalar potential is found. In particular, the points on the left of
(and above) the allowed region are ruled out because the minimum with
$\vev{H_u} \neq 0$ and $\vev{H_d} = \vev{N} =0$ is deeper than the
correct vacuum, see eq.~(\ref{wrongvac}). The points on the right of
the allowed region for $\lambda(M_S) \,\lsim \,0.5$ are ruled out by
the requirement that the top Yukawa coupling be perturbative up to the
GUT scale. For $\lambda(M_S) \,\gsim\, 0.5$ we find $\tan\beta
>1.5\,$, resulting in a not-too-large top Yukawa coupling. Therefore,
the upper-right arm of the allowed region can extend up to $\xi_U \sim
1.1$, where the couplings $\lambda$ and $k$ approach the
perturbativity bound at the GUT scale.

For values of $\xi_U$ in the vicinity of the right edge of the allowed
region $\tan\beta$ goes down to about 1.4. When $\xi_U$ decreases,
$\tan\beta$ increases, reaching values greater than 10 in the leftmost
corner of the allowed region. In addition, values of $\tan\beta$
greater than 10 are obtained for large $\xi_U$ and small
$\lambda$. This behaviour can be qualitatively understood by
considering that eqs.~(\ref{min2}) and (\ref{muBmu})--(\ref{trovmu}),
in the limit of large $\tan\beta$, reduce to
\beq
\frac{1}{\tan\beta} \simeq \frac{k}{\lambda} 
\left( 1-\frac{A_\lambda}{A_kw}\right)~.
\label{bterm}
\eeq
For small $\lambda$ and small $\xi_{U}$, $k/\lambda$ at the weak scale
is determined by the critical line in \eq{critline} to be small (and
$w\gg 1/3$). On the other hand, for large $\xi_{U}$ we approach the
condition that the vacuum with non-vanishing $\vev{N}$ is nearly
degenerate with the origin, {\it i.e.} $w\simeq 1/3$. Since the
boundary condition at the messenger scale in \eq{aterm} gives
$A_k=3A_\lambda$, there is an approximate cancellation in \eq{bterm}.

The values of $\mu$ that result from the minimization of the scalar
potential are in general of the order of the stop masses, but they are
inversely correlated to the values of $\tan\beta$ shown in
fig.~\ref{fig-laxitb}. In the region where $\tan\beta$ is small $\mu$
gets as large as 2.8 TeV, while in the regions where $\tan\beta$ is
large $\mu$ goes down to 1.4 TeV.  Indeed, one can see from
eq.~(\ref{min1}) that a value of $\tan\beta$ close to 1 enhances
$\mu$, both because of the factor $\tan^2\beta -1$ in the denominator
and because a smaller $\tan\beta$ results in a larger $h_t$, enhancing
the stop contribution to the running of $\tilde m^2_{H_u}$.

Figure \ref{fig-laxi} shows the mass of the lightest CP-even Higgs
boson $h_1$ over the plane $\xi_U-\lambda(M_S)$, with the same inputs
as in fig.~\ref{fig-laxitb}.  It can be seen from fig.~\ref{fig-laxi}
that there are three separate regions in the plane
$\xi_U-\lambda(M_S)$ where the mass of the lightest CP-even Higgs
boson is sufficiently large. These three regions can be characterized
as follows:
\paragraph{Region I.}
In the lower-left corner of the plot, where both $\lambda$ and
$\xi_U\,\ll 1$, the singlet vev $\vev{N}$ is generated by a large and
negative value of ${\tilde m}_N^2$, and $\tan\beta$ is proportional to
$\lambda/k$, taking on relatively large values $4 \lsim \tan\beta
\lsim 20$.  The tree-level mass of the lightest Higgs boson is
dominated by the first term in the r.h.s.~of eq.~(\ref{massh1}), as in
the MSSM, and the radiative corrections lift $m_{h_1}$ above the LEP
bound. However, for extremely small values of $\xi_U$ the ratio
$\lambda/k$ becomes very large, and the term unsuppressed by
$\sin2\beta$ in the curly brackets of eq.~(\ref{massh1}) gives a large
and negative contribution to $m_{h_1}$, dragging it again below the
LEP bound.  This is one of the reasons why we cannot consider the
N-GMSB with $\xi_U=0$.  The masses of the MSSM-like heavy Higgs bosons
$h_2$ and $a_1$ are of the order of $\mu$, while $h_3$ and $a_2$
become much lighter as $\tan\beta$ grows, since $m_{h_3} \sim
2\,(k/\lambda)\,\mu$ and $m_{a_2} \sim \sqrt{3/w}\,(k/\lambda)\,\mu$.
Note that $a_2$ becomes an approximate $R$-axion, because $A_k \to 0$.
The fermionic component of $N$ (singlino) has mass $M_{\widetilde N}
\sim m_{h_3}$. Since in this region $k/\lambda$ is small, the
singlet-like scalars and the singlino are considerably lighter than
the other non-SM particles, and the singlino can be the NLSP.  The
light pseudoscalar has sufficiently small couplings to escape LEP
bounds. The NNLSP is a bino-like neutralino, which decays into the
singlino and a (real or virtual) SM-like Higgs boson $h_1$. In this
scenario, the decay chains of supersymmetric particles end with the
NLSP singlino decaying into a pseudoscalar singlet and a gravitino,
with a rate
\beq
\Gamma\,(\widetilde N \rightarrow a_2\,\widetilde G) ~=~ 
\frac{\left(M_{\widetilde N}^2-m_{a_2}^2\right)^4}
{16\,\pi\,M_{\widetilde N}^3\,F^2}~.
\eeq
This process is a peculiar characteristic of the N-GMSB
model. However, the NLSP decay can occur inside the detector only for
$\sqrt{F}$ roughly smaller than $10^6$~GeV, a region disfavored by the
LEP bound on $m_{h_1}$.

\paragraph{Region II.}
In the lower-right corner of the plot, where $\lambda$ is small but
$\xi_U$ is large, the soft mass ${\tilde m}_N^2$ is positive, and the
vev $\vev{N}$ is generated by the large value of $A_k^2$, see
eqs.~(\ref{aterm}) and (\ref{spot}). The parameter $\tan\beta$ is
large due to the cancellation in eq.~(\ref{bterm}) obtained for
$w\simeq 1/3$, and, in contrast with what happens in region I, the
ratio $k/\lambda$ is large.  As a result, the negative and
$\tan\beta$-unsuppressed contributions to $m_{h_1}$ in
eq.~(\ref{massh1}) are not important, and the tree-level mass of the
lightest Higgs boson is approximately equal to $M_Z$ (as in the MSSM).
The other particles have masses $m_{a_2}/3 \sim M_{\widetilde N}/2
\sim m_{h_3} \sim (k/\lambda)\,\mu$, while $h_2$ and $a_1$ have masses
of order $\mu$.  Due to the large value of $k/\lambda$ the
singlet-like scalars and the singlino are much heavier than the other
scalars and neutralinos, making the particle spectrum similar to the
one of ordinary gauge mediation.

\paragraph{Region III.}
The last region with relatively large $m_{h_1}$ lies at large values
of $\lambda$ and close to the right edge of the region allowed by
perturbativity of the couplings, where $\tan\beta<2$. The soft
parameters $\tilde m_N^2$ and $A_k$ are large and negative, and they
both contribute to generating $\vev{N}$. The ratio $k/\lambda$ is
close to 1, therefore all the heavy scalars, as well as the higgsinos
and the singlino, have masses of the order of $\mu$, while the NLSP is
the bino-like neutralino. Concerning the mass of the lightest Higgs
boson $h_1$, the first term in the r.h.s.~of eq.~(\ref{massh1}) is
suppressed by the low value of $\tan\beta$, but the term
$\lambda^2\,v^2\sin^22\beta$ is sizeable and lifts $m_{h_1}$ above the
LEP bound.

If we give up the requirement that the couplings be perturbative up to
the GUT scale, considering only their evolution up to a relatively
small messenger scale, we can accommodate larger values of
$\lambda(M_S)$, resulting in a larger tree-level contribution to
$m_{h_1}$. For example, for $M = 10^7$ GeV, $F/M = 1.5 \times 10^5$
GeV (so that the stop masses are of the order of 2 TeV) and
$\lambda(M_S) \sim 0.7$ we can find a range of values of
$\xi_{D,T}(M_S)$ for which $m_{h_1} \sim 150$ GeV.

\begin{figure}[t]
\begin{center}
{\epsfig{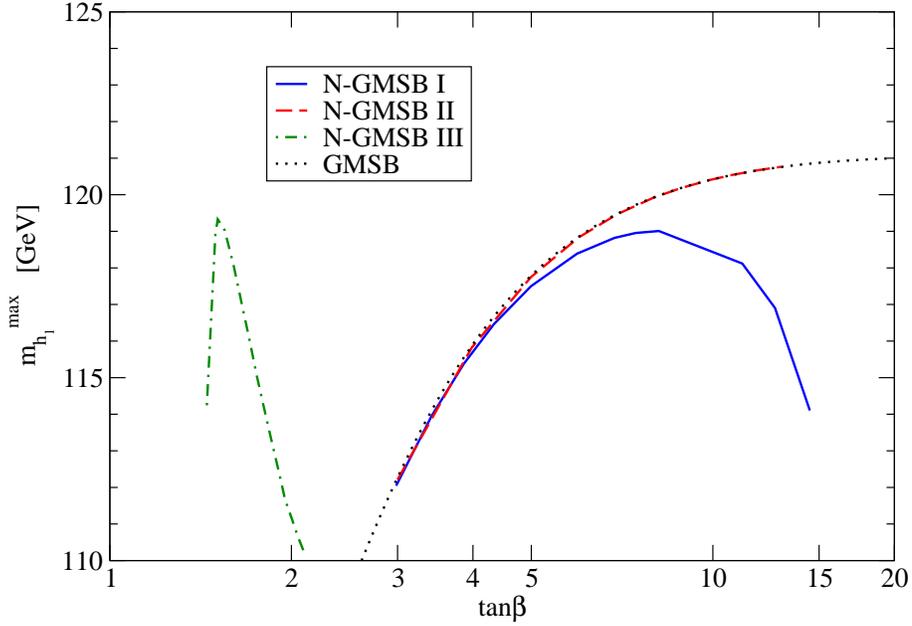}}
\end{center}
\caption{Upper bound on the mass of the lightest CP-even Higgs boson 
$h_1$ in N-GMSB and GMSB as a function
of $\tan\beta$, for $M = 10^{13}$ GeV and $F/M = 1.72
\times 10^5$ GeV.}
\label{fig-mhtb}
\end{figure}

It is interesting to compare the results for $m_{h_1}$ obtained in the
N-GMSB with those that are obtained in the usual GMSB for the same
values of $M$, $F/M$, and $\tan\beta$. The result is summarized in
fig.~\ref{fig-mhtb}, where we show the maximal value of $m_{h_1}$ as a
function of $\tan\beta$ for the GMSB and for the three
phenomenologically viable regions of the N-GMSB. In region III, for
low $\tan\beta$, the Higgs mass can be larger than the corresponding
value in GMSB. Region II gives a prediction for the maximum value of
$m_{h_1}$ that is identical to the one of GMSB. Finally, the upper
bound on $m_{h_1}$ in region I is approximately equal to the one of
GMSB for intermediate values of $\tan\beta$, but becomes smaller at
large $\tan\beta$, because of the negative contribution $-\lambda^4
v^2/k^2$ in \eq{massh1}.

\section{Conclusions}

There are two aspects of the $\mu$ problem. The first is related to
the absence of a $\mu$ term in the limit of exact supersymmetry and to
its generation from supersymmetry breaking. The second aspect (the
``$B_\mu$ problem") is related to the generic expectation $B_\mu /\mu
\simeq F/M$, and it is present only in models where the soft terms are
calculable and turn out parametrically smaller than the original seed
of supersymmetry breaking $F/M$. Gauge mediation belongs to this class
of models. For these models, as opposed to the usual supergravity
scenarios (where there is no $B_\mu$ problem), the extension of the
theory by adding a weak-scale singlet $N$ is more justifiable, because
it circumvents the unwanted relation $B_\mu /\mu \simeq F/M$, it does
not add many new free parameters, and it cannot destabilize the
hierarchy, at least for sufficiently low mediation scale.

In this paper, we have studied a model of gauge-mediation with the
addition of a singlet, including a messenger-singlet coupling, that
was first proposed in ref.~\cite{wave} (we denote this model as
N-GMSB). We have computed the induced soft terms and analyzed the
phenomenological consequences.  The singlet interactions are described
by the three couplings $\lambda$, $k$ and $\xi_U$. However, since two
of them can be traded for $\mu$ and $B_\mu$, the theory contains only
one extra parameter with respect to the ordinary gauge-mediated
supersymmetric SM without singlet, and therefore maintains its high
predictive power. Electroweak breaking requires a mild fine tuning,
endemic to supersymmetric models after LEP2. However, the necessary
``critical" condition can be achieved for a large range of values of
the coupling constants $\lambda$, $k$ and $\xi_U$ (at the price of
tuning one of the three parameters).

The lightest Higgs mass gives the most stringent constraint to N-GMSB,
leading to a heavy supersymmetric mass spectrum and a large messenger
scale $M$, and identifying three special regions in the space of
couplings $\lambda$, $k$ and $\xi_U$. Region I is characterized by
large $\tan\beta$ and light singlet fields. The singlet pseudoscalar
is particularly light and the singlino can be the NLSP, leading to a
potentially characteristic signal of the supersymmetric decay chains,
with Higgs and missing energy in the final states. Region II has large
$\tan\beta$, heavy singlet fields and a low-energy mass spectrum that
is very similar to the one of usual gauge mediation. Region III has
$\tan\beta$ close to one, and the tree-level quartic Higgs coupling is
mostly generated by $\lambda$. Anyway, even in these regions the
lightest Higgs mass is not larger than the maximal value attainable in
the usual GMSB. We find that the Higgs can be substantially heavier
than 120 GeV only if we abandon perturbativity of the couplings up to
the GUT scale.

\section*{Acknowledgments}
We thank Z.~Chacko, M.~Dine, J.~Mason and R.~Rattazzi for useful
discussions.

\newpage

\section*{Appendix A}

In this appendix we provide the RG equations for the gauge and
superpotential couplings of N-GMSB, valid above the messenger scale
$M$. The RG equations for the couplings and the soft
supersymmetry-breaking terms of the N-MSSM, valid below the scale $M$,
can be found e.g.~in ref.~\cite{muray} (note however that our
definition of $A_\lambda$ differs by a sign from that of
ref.~\cite{muray}).

Defining $\beta_{\lambda_i}$ as in eq.~(\ref{betarun}), the RG
equations for the couplings $\lambda_i$ are:
\bea
\beta_{g^\prime} &=& \frac{{g^{\prime\,}}^4}{8\pi^2}\,
\left(\frac{5\,n}{3}+11\right)~,\\
\beta_g &=& \frac{g^4}{8\pi^2}\,(n+1)~,\\
\beta_{g_s} &=& \frac{g_s^4}{8\pi^2}\,(n-3)~,\\
\beta_{h_t} &=& \frac{h_t^2}{8\pi^2}\,
\biggr(6\,h_t^2 +h_b^2 +\lambda^2-3\, g^2- \frac{13}{9} \,g^{\prime 2}
- \frac{16}{3}\, g_s^2\biggr)~,\\
\beta_{h_b} &=& \frac{h_b^2}{8\pi^2}\,
\biggr(6\,h_b^2 +h_t^2 +h_\tau^2+\lambda^2
-3\, g^2- \frac{7}{9} \,g^{\prime 2}- \frac{16}{3}\, g_s^2\biggr)~,\\
\beta_{h_\tau} &=& \frac{h_\tau^2}{8\pi^2}\,
\biggr(4\,h_\tau^2 + 3\,h_b^2+\lambda^2
-3\, g^2 - 3\,g^{\prime 2}\biggr)~,\\
\beta_{\lambda} &=& \frac{\lambda^2}{8\pi^2}\,
\biggr(4\,\lambda^2+2\,k^2 + 3\,h_t^2 + 3\,h_b^2 +h_\tau^2
+2\,\xi_D^2+3\,\xi_T^2 - 3\, g^2 - g^{\prime 2}\biggr)~,\\
\beta_{k} &=& \frac{k^2}{8\pi^2}\,
\biggr(6\,\lambda^2+6\,k^2 + 6 \,\xi_D^2 + 9\,\xi_T^2\biggr)~,\\
\beta_{\xi_D}&=&\frac{\xi_D^2}{8\pi^2}\,
\biggr( 2\,\lambda^2 +2\,k^2 +4 \,\xi_D^2+3\,\xi_T^2
-3\,g^2-g^{\prime 2}\biggr)~,\\
\beta_{\xi_T}& =&\frac{\xi_T^2}{8\pi^2}\,
\left( 2\,\lambda^2 +2\,k^2 +2 \,\xi_D^2+5\,\xi_T^2-\frac{16}{3} 
\,g_s^2-\frac 49 \,g^{\prime 2}\right) ~,
\eea
where $n$ is the number of messenger pairs.

\newpage

\section*{Appendix B}

In this appendix we derive the expressions for the soft supersymmetry
breaking terms in the scalar sector using the wave-function
renormalization method proposed in refs.~\cite{wave,wave2}.

The soft terms in eqs.~(\ref{sf1})--(\ref{sf2}) are given in terms of
the following derivatives of the wave-function renormalization $Z$
with respect to the messenger mass $M$, evaluated at the
renormalization scale $Q$ (for simplicity we will drop the field index
$\alpha$ in this appendix)
\beq 
{\cal Z}^{\prime} 
= \frac{\partial \ln Z \left( M,Q\right)}{2~\partial \ln M},
~~~~ {\cal Z}^{\prime\prime} 
= \frac{\partial^2 \ln Z \left( M,Q\right)}{4~\partial (\ln M)^2 }.  
\eeq

By integrating \eq{zrun} between an arbitrary high-energy scale
$\Lambda$ and the renormalization scale $Q$ (with $Q<M$), we obtain
\beq
\ln \frac{Z(Q)}{Z(\Lambda)} = \int_{\ln \Lambda}^{\ln M} dt~ 
\gamma^{(+)} + \int_{\ln M}^{\ln Q} dt ~\gamma^{(-)}  ,
\label{intan}
\eeq
where $\gamma^{(\pm)}$ are the anomalous dimensions above and below
the messenger scale $M$, respectively.

Taking the first derivative of \eq{intan}, we obtain the expression of
${\cal Z}^{\prime}(Q=M)$ shown in \eq{derivv}. Taking the second
derivative, we find
\beq
\left. {\cal Z}^{\prime\prime} \right|_{Q=M} = 
\frac 14 \sum_i \left[ \frac{\partial \left( \Delta \gamma \right)}
{\partial \lambda_i^2} \frac{\partial\lambda_i^2(M)}
{\partial \ln M} -\frac{\partial \gamma^{(-)}}
{\partial \lambda_i^2} \left. \frac{\partial\lambda_i^2(Q)}
{\partial \ln M}\right|_{Q=M} \right] .
\label{zzz}
\eeq
With the help of \eq{betarun}, we obtain
\beq
\frac{\partial\lambda_i^2(M)}{\partial \ln M}= 
\beta_{\lambda_i}^{(+)}\big |_{Q=M} ,~~~
\left. \frac{\partial\lambda_i^2(Q)}{\partial \ln M}\right|_{Q=M} 
= \Delta \beta_{\lambda_i} .
\label{soup}
\eeq
Replacing \eq{soup} into \eq{zzz}, we obtain the expression of ${\cal
Z}^{\prime\prime}(Q=M)$ shown in \eq{derivv}.


\newpage

\begin{thebibliography}{99}

\bibitem{dine1}
  M.~Dine and A.~E.~Nelson,
  Phys.\ Rev.\  D {\bf 48} (1993) 1277
  [arXiv:hep-ph/9303230].
  
\bibitem{dine2}
  M.~Dine, A.~E.~Nelson and Y.~Shirman,
  Phys.\ Rev.\  D {\bf 51} (1995) 1362
  [arXiv:hep-ph/9408384].

\bibitem{dine3}
  M.~Dine, A.~E.~Nelson, Y.~Nir and Y.~Shirman,
  Phys.\ Rev.\  D {\bf 53} (1996) 2658
  [arXiv:hep-ph/9507378].
  
\bibitem{grrev}
  G.~F.~Giudice and R.~Rattazzi,
  Phys.\ Rept.\  {\bf 322} (1999) 419
  [arXiv:hep-ph/9801271].

\bibitem{meta}
  S.~Dimopoulos, G.~R.~Dvali, R.~Rattazzi and G.~F.~Giudice,
  Nucl.\ Phys.\  B {\bf 510} (1998) 12
  [arXiv:hep-ph/9705307].

  \bibitem{meta2}
    M.~A.~Luty,
  Phys.\ Lett.\  B {\bf 414} (1997) 71
  [arXiv:hep-ph/9706554];
    S.~Dimopoulos, G.~R.~Dvali and R.~Rattazzi,
  Phys.\ Lett.\  B {\bf 413} (1997) 336
  [arXiv:hep-ph/9707537].
  
\bibitem{iss}
  K.~Intriligator, N.~Seiberg and D.~Shih,
  JHEP {\bf 0604} (2006) 021
  [arXiv:hep-th/0602239].

\bibitem{issmod}
  S.~Franco and A.~M.~Uranga,
  JHEP {\bf 0606} (2006) 031
  [arXiv:hep-th/0604136];
 H.~Ooguri and Y.~Ookouchi,
  Nucl.\ Phys.\  B {\bf 755} (2006) 239
  [arXiv:hep-th/0606061];
    T.~Banks,
  arXiv:hep-ph/0606313;
    R.~Kitano,
  Phys.\ Lett.\  B {\bf 641} (2006) 203
  [arXiv:hep-ph/0607090];
   S.~Forste,
  Phys.\ Lett.\  B {\bf 642} (2006) 142
  [arXiv:hep-th/0608036];
    M.~Dine, J.~L.~Feng and E.~Silverstein,
  Phys.\ Rev.\  D {\bf 74} (2006) 095012
  [arXiv:hep-th/0608159];
     M.~Dine and J.~Mason,
  arXiv:hep-ph/0611312;
  R.~Kitano, H.~Ooguri and Y.~Ookouchi,
  Phys.\ Rev.\  D {\bf 75} (2007) 045022
  [arXiv:hep-ph/0612139];
   H.~Murayama and Y.~Nomura,
  Phys.\ Rev.\ Lett.\  {\bf 98} (2007) 151803
  [arXiv:hep-ph/0612186];
    C.~Csaki, Y.~Shirman and J.~Terning,
  arXiv:hep-ph/0612241;
   O.~Aharony and N.~Seiberg,
  JHEP {\bf 0702} (2007) 054
  [arXiv:hep-ph/0612308];
    S.~A.~Abel and V.~V.~Khoze,
  arXiv:hep-ph/0701069;
  H.~Murayama and Y.~Nomura,
  Phys.\ Rev.\  D {\bf 75} (2007) 095011
  [arXiv:hep-ph/0701231].

\bibitem{giumas}
  G.~F.~Giudice and A.~Masiero,
  Phys.\ Lett.\  B {\bf 206} (1988) 480.

\bibitem{anom}
  L.~Randall and R.~Sundrum,
  Nucl.\ Phys.\  B {\bf 557} (1999) 79
  [arXiv:hep-th/9810155];
  G.~F.~Giudice, M.~A.~Luty, H.~Murayama and R.~Rattazzi,
  JHEP {\bf 9812} (1998) 027
  [arXiv:hep-ph/9810442].
  
\bibitem{gaug}
  D.~E.~Kaplan, G.~D.~Kribs and M.~Schmaltz,
  Phys.\ Rev.\  D {\bf 62} (2000) 035010
  [arXiv:hep-ph/9911293];
  Z.~Chacko, M.~A.~Luty, A.~E.~Nelson and E.~Ponton,
  JHEP {\bf 0001} (2000) 003
  [arXiv:hep-ph/9911323].

\bibitem{dynmu}
 G.~R.~Dvali, G.~F.~Giudice and A.~Pomarol,
  Nucl.\ Phys.\  B {\bf 478} (1996) 31
  [arXiv:hep-ph/9603238].

\bibitem{scamu}
  K.~Choi and H.~D.~Kim,
  Phys.\ Rev.\  D {\bf 61} (2000) 015010
  [arXiv:hep-ph/9906363];
  M.~Ito,
  Prog.\ Theor.\ Phys.\  {\bf 106} (2001) 577
  [arXiv:hep-ph/0011004];
  R.~Kitano,
  Phys.\ Rev.\  D {\bf 74} (2006) 115002
  [arXiv:hep-ph/0606129].
  
\bibitem{cosmu}
  T.~Yanagida,
  Phys.\ Lett.\  B {\bf 400} (1997) 109
  [arXiv:hep-ph/9701394].
  
\bibitem{flamu}
  K.~S.~Babu and Y.~Mimura,
  arXiv:hep-ph/0101046.

\bibitem{errmu}
  L.~J.~Hall, Y.~Nomura and A.~Pierce,
  Phys.\ Lett.\  B {\bf 538} (2002) 359
  [arXiv:hep-ph/0204062].

\bibitem{destab}
  J.~Polchinski and L.~Susskind,
  Phys.\ Rev.\  D {\bf 26} (1982) 3661;
  H.~P.~Nilles, M.~Srednicki and D.~Wyler,
  Phys.\ Lett.\  B {\bf 124} (1983) 337.

\bibitem{stab}
  D.~Nemeschansky,
  Nucl.\ Phys.\  B {\bf 234} (1984) 379.

\bibitem{slid}
  E.~Witten,
  Phys.\ Lett.\  B {\bf 105} (1981) 267.
  
\bibitem{pomar}
  P.~Ciafaloni and A.~Pomarol,
  Phys.\ Lett.\  B {\bf 404} (1997) 83
  [arXiv:hep-ph/9702410].

\bibitem{nir}
  P.~Langacker, N.~Polonsky and J.~Wang,
  Phys.\ Rev.\  D {\bf 60} (1999) 115005
  [arXiv:hep-ph/9905252].

\bibitem{muray}
  A.~de Gouvea, A.~Friedland and H.~Murayama,
  Phys.\ Rev.\  D {\bf 57} (1998) 5676
  [arXiv:hep-ph/9711264].

\bibitem{wave}
  G.~F.~Giudice and R.~Rattazzi,
  Nucl.\ Phys.\  B {\bf 511} (1998) 25
  [arXiv:hep-ph/9706540].
  
\bibitem{han}
  T.~Han, D.~Marfatia and R.~J.~Zhang,
  Phys.\ Rev.\  D {\bf 61} (2000) 013007
  [arXiv:hep-ph/9906508].

\bibitem{chacko}
  Z.~Chacko and E.~Ponton,
  Phys.\ Rev.\  D {\bf 66} (2002) 095004
  [arXiv:hep-ph/0112190];
%
  Z.~Chacko, E.~Katz and E.~Perazzi,
  Phys.\ Rev.\  D {\bf 66} (2002) 095012
  [arXiv:hep-ph/0203080].

\bibitem{wave2}
 N.~Arkani-Hamed, G.~F.~Giudice, M.~A.~Luty and R.~Rattazzi,
  Phys.\ Rev.\  D {\bf 58} (1998) 115005
  [arXiv:hep-ph/9803290].

\bibitem{mtop}
  Tevatron Electroweak Working Group for the CDF and D0 Collaborations,
  [arXiv:hep-ex/0703034].

\bibitem{pdg}
  W.~M.~Yao {\it et al.}  [Particle Data Group],
  J.\ Phys.\ G {\bf 33} (2006) 1.

\bibitem{schmaltz}
  A.~G.~Cohen, T.~S.~Roy and M.~Schmaltz,
  JHEP {\bf 0702} (2007) 027
  [arXiv:hep-ph/0612100].

\bibitem{crit}
  G.~F.~Giudice and R.~Rattazzi,
  Nucl.\ Phys.\  B {\bf 757} (2006) 19
  [arXiv:hep-ph/0606105].

\bibitem{nmssmhiggs}
  J.~R.~Ellis, J.~F.~Gunion, H.~E.~Haber, L.~Roszkowski and F.~Zwirner,
  Phys.\ Rev.\  D {\bf 39} (1989) 844;
%
  M.~Drees,
  Int.\ J.\ Mod.\ Phys.\  A {\bf 4} (1989) 3635;
%
   U.~Ellwanger, M.~Rausch de Traubenberg and C.~A.~Savoy,
  Phys.\ Lett.\  B {\bf 315} (1993) 331
  [arXiv:hep-ph/9307322],
%
  Nucl.\ Phys.\  B {\bf 492} (1997) 21
  [arXiv:hep-ph/9611251];
%
  S.~F.~King and P.~L.~White,
  Phys.\ Rev.\  D {\bf 52} (1995) 4183
  [arXiv:hep-ph/9505326];
 
\bibitem{mssmcorr}
  J.~R.~Ellis, G.~Ridolfi and F.~Zwirner,
  Phys.\ Lett.\  B {\bf 257} (1991) 83;
%
  H.~E.~Haber and R.~Hempfling,
  Phys.\ Rev.\ Lett.\  {\bf 66} (1991) 1815;
%
  Y.~Okada, M.~Yamaguchi and T.~Yanagida,
  Prog.\ Theor.\ Phys.\  {\bf 85} (1991) 1.

\bibitem{LEPhiggs}
  R.~Barate {\it et al.}  [LEP Working Group for Higgs boson searches],
  Phys.\ Lett.\  B {\bf 565} (2003) 61
  [arXiv:hep-ex/0306033];
%
  S.~Schael {\it et al.}  [ALEPH Collaboration],
  Eur.\ Phys.\ J.\  C {\bf 47} (2006) 547
  [arXiv:hep-ex/0602042].

\bibitem{nmssmcorr}
  U.~Ellwanger,
  Phys.\ Lett.\  B {\bf 303} (1993) 271
  [arXiv:hep-ph/9302224];
%
  T.~Elliott, S.~F.~King and P.~L.~White,
  Phys.\ Lett.\  B {\bf 305} (1993) 71
  [arXiv:hep-ph/9302202],
%
  Phys.\ Lett.\  B {\bf 314} (1993) 56
  [arXiv:hep-ph/9305282],
%
  Phys.\ Rev.\  D {\bf 49} (1994) 2435
  [arXiv:hep-ph/9308309];
%
  P.~N.~Pandita,
  Phys.\ Lett.\  B {\bf 318} (1993) 338,
%
  Z.\ Phys.\  C {\bf 59} (1993) 575.

\bibitem{nmhdecay}
  U.~Ellwanger, J.~F.~Gunion and C.~Hugonie,
  JHEP {\bf 0502} (2005) 066
  [arXiv:hep-ph/0406215];
%
  U.~Ellwanger and C.~Hugonie,
  Comput.\ Phys.\ Commun.\  {\bf 175} (2006) 290
  [arXiv:hep-ph/0508022],
%
  arXiv:hep-ph/0612134.

\end{thebibliography}
\end{document}